\documentclass[journal]{IEEEtran}
\usepackage{amsmath,amsfonts}
\usepackage{algorithmic}
\usepackage{algorithm}
\usepackage{array}
\usepackage[caption=false,font=normalsize,labelfont=sf,textfont=sf]{subfig}
\usepackage{textcomp}
\usepackage{stfloats}
\usepackage{url}
\usepackage{verbatim}
\usepackage{graphicx}
\usepackage{setspace}
\usepackage{cite}
\usepackage{float}
\usepackage{amssymb}
\renewcommand{\algorithmicrequire}{\textbf{Input:}}  
\usepackage{booktabs}
\usepackage{color}

\makeatletter
\renewcommand{\maketag@@@}[1]{\hbox{\m@th\normalsize\normalfont#1}}%
\makeatother

\begin{document}
	\vspace{-0.1cm}
	\title{\LARGE{Rotatable Antenna-Enhanced Cell-Free Communication}}
	
	\author{Kecheng Pan, Beixiong Zheng, \textit{Senior Member, IEEE,} Yanhua Tan, Fangjiong Chen, \textit{Member, IEEE}, Emil Björnson,  \textit{Fellow, IEEE,} Robert~Schober, \textit{Fellow, IEEE,} and Rui Zhang, \textit{Fellow, IEEE} \IEEEmembership{}
		\vspace{-0.4cm}	
		\thanks{The work of Fangjiong Chen was supported by National Key Research and Development Program of China under Grant 2025YFE0214800. The work of Beixiong Zheng was supported in part by the National Natural Science Foundation of China under Grant 62571193 and Grant 62331022, the Guangdong program under Grant 2023QN10X446 and Grant 2023ZT10X148, and the GJYC program of Guangzhou under Grant 2024D01J0079 and Grant 2024D03J0006. The work of Robert Schober was supported in part by the Federal Ministry for Research, Technology and Space (BMFTR) in Germany in the program of ``Souverän. Digital. Vernetzt." joint project xG-RIC under Project 16KIS2432. (\textit{Corresponding author: Beixiong Zheng.)}}
		\thanks{Kecheng Pan, Beixiong Zheng, and Yanhua Tan are with the School of Microelectronics, South China University of Technology, Guangzhou 511442, China (e-mails: 202421065214@mail.scut.edu.cn; bxzheng@scut.edu.cn; tanyanhua06@163.com).}
		\thanks{Fangjiong Chen, is with School of Electronic and Information Engineering, South China University of Technology, Guangzhou, 510640 China, (email:eefjchen@scut.edu.cn;).}
		\thanks{Emil Björnson is with the Division of Communication Systems, KTH Royal Institute of Technology, 100 44 Stockholm, Sweden (e-mail:  emilbjo@kth.se).}
		\thanks{Robert Schober is with the Institute for Digital Communications, Friedrich Alexander-University Erlangen-Nürnberg, 91054 Erlangen, Germany (e-mail: robert.schober@fau.de).}
		\thanks{Rui Zhang is with the Department of Electrical and Computer Engineering, National University of Singapore, Singapore 117583 (e-mail: elezhang@nus.edu.sg).}}

	
	
	\maketitle
	
	\begin{abstract}
		Rotatable antenna (RA) is a promising technology that can exploit new design dimensions by optimizing the three-dimensional (3D) boresight directions of directional antennas. In this letter, we investigate an RA-enhanced cell-free system for downlink transmission, where multiple RA-equipped access points (APs) cooperatively serve multiple single-antenna users over the same time-frequency resource. Specifically, we aim to maximize the sum rate of all users by jointly optimizing the AP-user associations and the RA boresight directions. Accordingly, we propose a two-stage strategy to solve the AP-user association problem, and then employ fractional programming (FP) and successive convex approximation (SCA) techniques to optimize the RA boresight directions. Numerical results demonstrate that the proposed RA-enhanced cell-free system significantly outperforms various benchmark schemes. 
	\end{abstract}
	
	\vspace{-0.5cm}
	\section{Introduction}
	With the rapid development of global information and communication technology (ICT), the denser deployment of access points (APs) has become a prevailing trend in communication systems to meet the increasing demands for lower latency, higher reliability, and improved spectral efficiency. A promising paradigm is the cell-free multiple-input multiple-output (MIMO) system \cite{ref1}, where numerous APs, distributed throughout the service area and connected to a central processing unit (CPU) via fronthaul, collaboratively serve all users and thereby alleviate cell boundaries \cite{ref2}. Conventional cell-free MIMO systems typically employ isotropic and fixed antennas, whose radiation directions cannot be adjusted after deployment, thereby restricting performance. 
	
	To address these limitations, rotatable antenna (RA) has recently emerged as a promising solution, enabling a 34
	new design dimension by dynamically adjusting the three-dimensional (3D) boresight direction of antennas. By the 
	rotation flexibility (without changing antenna positions), RA enables a cost-effective and compact design more suitable for practical deployment \cite{ref3,ref4}. Due to these attractive advantages, recent studies have explored the performance gains enabled by RA in various wireless systems. The authors in \cite{ref8} established the fundamental system model and channel characteristics of RA systems, demonstrating their ability to achieve more favorable channel conditions. Building on this, RA has been applied to more challenging scenarios, such as integrated sensing and communication (ISAC)\cite{ref9} and physical-layer security\cite{ref10}. Notably, by enabling flexible 3D boresight adaptation at distributed APs, RA can strengthen desired links while reducing interference toward unintended users. The main benefit of this design is that it can achieve spatial adaptation with only one antenna per AP, providing a lightweight and distributed implementation method. In contrast, traditional fixed arrays require more antennas to achieve similar directional capabilities. Nevertheless, the application of RA in such systems remains unexplored, thus motivating further investigation in this work.
	
	Motivated by the above, in this letter, we investigate an RA-enhanced cell-free MIMO system for downlink transmission (as shown in Fig. 1), where multiple RA-equipped APs collaboratively provide communication services to distributed users. To enhance the communication quality of users, we formulate an optimization problem for maximizing their sum rate by jointly designing the AP-user associations and the RA boresight directions. As the formulated problem is non-convex, we propose a two-stage algorithm to determine the AP-user associations, and then apply fractional programming (FP) with successive convex approximation (SCA) to produce high-quality solutions for the RA boresight directions. Simulation results show that RA can enhance the sum rate of all users, significantly outperforming various benchmark schemes.

	\begin{figure}[!t]
		\vspace{-0.3cm}
		\centering
		\includegraphics[width=0.8\linewidth]{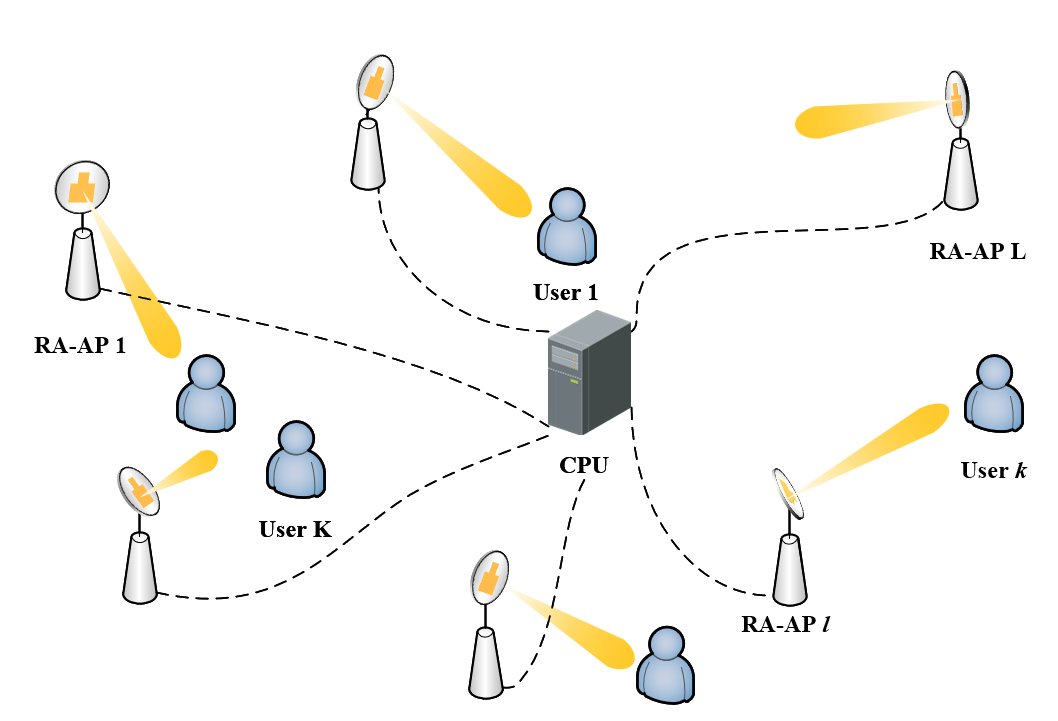}
		\caption{\centering{Cell-free system with RA-equipped APs.}}
		\label{fig_1}
		\vspace{-5mm}
	\end{figure}
	
	\vspace{-0.3cm}
	\section{System Model and Problem Formulation}
	As shown in Fig. 1, we consider an RA-enhanced downlink cell-free network consisting of $L$ APs and $K$ users, where $L\ge K$. Each AP is equipped with a directional RA, while each user employs a single isotropic antenna. All APs are connected to a CPU via a fronthaul network and cooperatively serve users over the same time-frequency resource. Let $\mathcal L =\left\{ 1,2,...,L \right\}$ and $ \mathcal K =\left\{ 1,2,...,K \right\} $ denote the index sets of the APs and users, respectively. Without loss of generality, we assume that all APs and users are located in a global Cartesian coordinate system (CCS). Let $\boldsymbol{v}_l\in \mathbb{R}^{3\times 1} ,l \in \mathcal L$, and $\boldsymbol{q}_k\in \mathbb{R}^{3\times 1},k \in \mathcal K$ denote the positions of AP $l$ and user $k$, respectively. Let $d_{l,k}=\lVert \boldsymbol{q}_k- \boldsymbol{v}_l \rVert $ denote the distance between AP $l$ and user~$k$, where $\lVert \cdot \rVert$ is the Euclidean norm.
	\vspace{-4mm}
	\subsection{Channel Model}
	In this letter, we assume a generic directional gain pattern for each RA as \cite{ref7}
	
	\vspace{-0.3cm}
	\small
	\begin{equation}
		G_e\left( \epsilon ,\varphi \right) =\begin{cases}
			G_0\cos ^{2p}\left( \epsilon \right) ,&		\epsilon \in \left[ 0,\frac{\pi}{2} \right) ,\varphi \in \left[ 0,2\pi \right) \\ \label{1}
			0,&		\text{otherwise,}\\
		\end{cases}
	\end{equation}\normalsize
	where $\left( \epsilon ,\varphi \right)$ is a pair of incident angles with respect to the RA boresight direction, $p$ denotes the directivity factor, and $G_0 = 2(2p+1)$  denotes the maximum gain in the boresight direction. Let $\vec{\mathbf{f}}_l$ denote the RA pointing vector at AP $l$. Accordingly, the directional gain of the RA at AP~$l$ in the direction of user $k$ can be expressed as $G_{l,k}=G_0\cos ^{2p}\left( \epsilon _{l,k} \right) $, where $\cos \left( \epsilon _{l,k} \right) \triangleq \vec{\mathbf{f}}_{l}^{T}\vec{\boldsymbol{q}}_{l,k}$ is the projection between $\vec{\mathbf{f}}_{l}^{T}$ and $\vec{\boldsymbol{q}}_{l,k}$ with $\vec{\boldsymbol{q}}_{l,k}\triangleq \frac{\boldsymbol{q}_k-\boldsymbol{v}_l}{d_{l,k}}$ denoting the normalized direction vector from AP $l$ to user $k$.
	
	We assume that the channel from AP $l$ to user $k$ experiences quasi-static flat-fading, which can be modeled as
	
	\vspace{-0.2cm}
	\small
	\begin{equation} 
		h_{l,k}\left( \vec{\mathbf{f}}_l \right) =\sqrt{\beta \left( d_{l,k} \right)}g_{l,k},
		\vspace{-0.2cm}
	\end{equation}\normalsize
	where $\beta\left( d_{l,k} \right)$ denotes the large-scale channel power gain due to the distance-dependent path loss and shadowing, and $g_{l,k}$ models the small-scale channel fading. Furthermore, we assume that $g_{l,k}$ follows an independent Rician fading distribution, characterized by Rician factor $\kappa$ and given by
	
	\vspace{-0.3cm}
	\small
	\begin{equation} 
		g_{l,k}=\sqrt{\frac{\kappa}{\kappa+1}}\bar{g}_{l,k}+\sqrt{\frac{1}{\kappa+1}}\tilde{g}_{l,k},
	\end{equation}\normalsize 
	where $\bar{g}_{l,k}=\sqrt{G_{l,k}}e^{-j\frac{2\pi}{\lambda}d_{l,k}}$ denotes the line-of-sight (LoS) component of the channel, and $\tilde{g}_{l,k}\sim \mathcal{C}\mathcal{N}\left( 0,1 \right) $ denotes the non-LoS (NLoS) channel component modeled by Rayleigh fading, with $\mathcal{CN}(0,1)$ denoting a circularly symmetric complex Gaussian (CSCG) distribution with mean zero and unit covariance.
	
	\subsection{Signal Model}
	Since a single RA can only steer its antenna boresight toward one direction at a time, for simplicity, we assume that each AP steers its antenna towards a specific user. Moreover, we use the matrix $\mathbf{B} \in \mathbb{R}^{L\times K}$ to represent the AP-user associations, which can be written as
	
	\vspace{-0.2cm}
	\small
	\begin{equation} 
		\mathbf{B}=\left[ \begin{matrix}
			b_{1,1}&		\cdots&		b_{1,K}\\
			\vdots&		\ddots&		\vdots\\
			b_{L,1}&		\cdots&		b_{L,K}\\
		\end{matrix} \right] ,
	\end{equation}\normalsize
	where $b_{l,k}=\left\{ 0,1 \right\} ,\forall l\in \mathcal{L},\forall k\in \mathcal{K}$,
	is a binary indicator for the association between AP $l$ and user $k$, where $b_{l,k}=1$ indicates that user $k$ is served by AP $l$; otherwise $b_{l,k}=0$. 
	
	As a prerequisite for precoding and subsequent optimization, we assume that global channel state information (CSI) can be fully acquired by the CPU via the fronthaul\cite{ref11,ref12}.
	
	In addition, to achieve coherent superposition and maximize the received signal power, we adopt normalized conjugate beamforming with precoding factor $\phi _{l,k}^{*}\triangleq \frac{h_{l,k}^{*}}{\left| h_{l,k} \right|}$. Thus, the downlink transmitted signal $s_l$ at AP $l$ can be written as 
	
	\vspace{-0.2cm}
	\small
	\begin{equation}
		s_l=\sqrt{P}\sum_{k=1}^K{b_{l,k}\phi _{l,k}^{*}x_k},
	\end{equation}\normalsize
	where $P$ denotes the transmit power of the AP, $x_k$ represents the data signal intended for user $k$, with $\mathbb{E}\left\{ \left| x_k \right|^2 \right\} =1$. Consequently, the received signal at user $k$ can be expressed as
	
	\vspace{-0.5cm}
	\small
	\begin{align}
		y_k\left( \vec{\mathbf{f}}_l \right) &=\sum_{l=1}^L{h_{l,k}\left( \vec{\mathbf{f}}_l \right) s_l+n_k} \nonumber
		=\underbrace{\sqrt{P}\sum_{l=1}^L{h_{l,k}\left( \vec{\mathbf{f}}_l \right) b _{l,k}\phi _{l,k}^{*}x_k}} _{\text{Desired signal}}\\
		&+\underbrace{\sqrt{P}\sum_{i\ne k}{\sum_{l=1}^L{h_{l,k}\left( \vec{\mathbf{f}}_l \right) b_{l,i}\phi _{l,i}^{*}x_i}}}_{\text{Inter-user interference}}+n_k, \label{7}
	\end{align}\normalsize
	where $n_k\thicksim \mathcal{CN} \left( 0,\sigma ^2 \right)$ represents complex additive white Gaussian noise (AWGN) with variance $\sigma ^2$. Thus, the received signal-to-interference-plus-noise ratio (SINR) at user $k$ can be expressed as
	
	\vspace{-0.2cm}
	\small
	\begin{equation}
		\gamma _k=\frac{P\left| \sum_{l=1}^L{h_{l,k}\left( \vec{\mathbf{f}}_l \right) b_{l,k}\phi _{l,k}^{*}} \right|^2}{P\sum_{i\ne k}{\left| \sum_{l=1}^L{h_{l,k}\left( \vec{\mathbf{f}}_l \right) b_{l,i}\phi _{l,i}^{*}} \right|}^2+\sigma ^2}.
	\end{equation}\normalsize
	Accordingly, the achievable rate of user $k$ can be expressed as
	
	\vspace{-0.2cm}
	\small
	\begin{equation}
		R_k=\log _2\left( 1+ \gamma_k\right) .
	\end{equation}\normalsize
	
	In this letter, we aim to jointly optimize the AP-user associations and the RA pointing vectors to maximize the sum rate of users. Accordingly, the optimization problem is formulated as
	
	\vspace{-0.5cm}
	\small
	\begin{subequations}\label{10}
		\begin{align}
			\left( \text{P}1 \right) :\max_{\mathcal{F,}\mathbf{B}}&\,\,\sum_{k=1}^K{R_k} \label{10a}\\
			\text{s.t.} \;&\lVert \vec{\mathbf{f}}_l \rVert =1,\forall l,\label{10b}\\
			&\sum_{l=1}^L{b_{l,k}\ge 1,\forall k\in \mathcal{K}},\label{10c}\\
			&\sum_{k=1}^K{b_{l,k}=1,\forall l\in \mathcal{L}},\label{10d}
		\end{align}
	\end{subequations}\normalsize
	where $\mathcal{F}=\left\{ \vec{\mathbf{f}}_1,...,\vec{\mathbf{f}}_L \right\} $ denotes the collection of RA pointing vectors. Constraint \eqref{10b} enforces the unit-norm condition for each pointing vector, \eqref{10c} ensures that each user is served by at least one AP for fairness, and \eqref{10d} ensures that each AP only provides services to one user. Note that (P1) is challenging to solve due to the non-concave objective \eqref{10a} and the non-convex constraint \eqref{10b}. To efficiently address the problem, we first propose a two-stage strategy to obtain the AP-user association matrix, and then develop a fractional programming-based algorithm to optimize the RA pointing vectors.

	\begin{figure}[t]
		\vspace{-0.0cm}
		\centering
		\includegraphics[width=\linewidth]{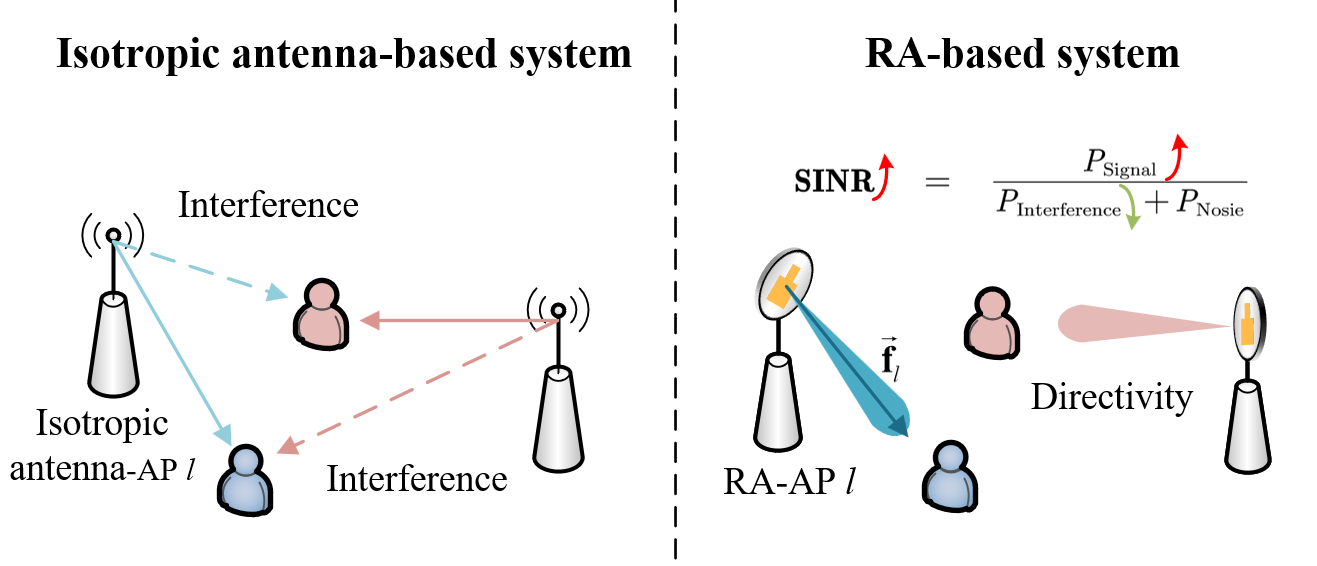}
		\caption{\centering{Comparison between an isotropic antenna-based system and an RA-based system.}}
		\label{fig_2}
		\vspace{-5mm}
	\end{figure}
	Before solving (P1),  we first illustrate the benefits of RA-based systems compared to conventional isotropic antenna-based systems in Fig. 2. In isotropic antenna-based systems, as the antennas’ broadcast signals uniformly in all directions, severe inter-user interference may occur. In contrast, in RA-based systems, the 3D boresight direction of each antenna can be adaptively adjusted toward the intended user, strengthening the desired signal while reducing interference to other users, thereby improving the overall SINR of users.
	
	\section{Proposed Algorithm}
	In this section, to tackle the non-convexity of Problem (P1), which involves integer variables and intricate coupling between the association matrix $\mathbf{B}$ and the RA pointing vectors $\mathcal{F}$, we decompose it into two sequential stages. We first develop a two-stage AP–user association strategy to obtain the association matrix $\mathbf{B}$. Subsequently, we propose a fractional programming-based algorithm to optimize the RA pointing vector at each AP, aiming to maximize the system sum rate.
	\vspace{-4mm}
	\subsection{AP-User Association Strategy}
	By exploiting the fact that pairing nearby APs and users causes less link attenuation and is more likely to achieve high performance, we approximate (P1) for AP-user pairing. Specifically, by calculating the distances $d_{l,k}$ of all possible AP-user pairs and ignoring interference, we reformulate the original association problem as minimizing the total AP-user association distance $D_{\text{sum}}=\sum_{l=1}^L{\sum_{k=1}^K{b_{l,k}d_{l,k}}}$. Accordingly, we have:
	
	\vspace{-0.4cm}
	\small
	\begin{subequations}
		\begin{align}
			\left( \text{P}2 \right) :\min\limits_{b_{l,k}}\ & D_{\text{sum}}\\
			\text{s.t.} \;&b_{l,k}\in \left\{ 0,1 \right\} ,\forall l\in \mathcal{L},\forall k\in \mathcal{K}, \label{11b} \\ 
			&\sum_{l=1}^L{b_{l,k}\ge 1,\forall k\in \mathcal{K},}\label{11c}\\
			&\sum_{k=1}^K{b_{l,k}=1,\forall l\in \mathcal{L}}\label{11d}.
		\end{align}
	\end{subequations}\normalsize
	Note that constraint \eqref{11b} defines the problem as a zero-one optimization problem, \eqref{11c} ensures that each user is associated with at least one AP, and \eqref{11d} ensures that each AP serves only one user. To address this non-convex problem, we propose a two-stage AP-user association strategy based on the AP-user distances. As summarized in Algorithm 1, the proposed strategy consists of two stages:
	
	\textit{1) Initial AP-User Association:} At this stage, we aim to allocate each user to a unique AP for fairness, while ensuring that the distance of each paired AP–user is as small as possible. To begin with, we initialize the index sets of unpaired APs and users as $\mathcal{L}^{\left( 0 \right)}=\mathcal L$ and $\mathcal{K}^{\left( 0 \right)}=\mathcal K$ with $\left| \mathcal{L}^{\left( 0 \right)} \right|=L$ and $\left| \mathcal{K}^{\left( 0 \right)} \right|=K$, and set the AP–user association matrix to $\mathbf{B}=\mathbf{0}_{L\times K}$, with $\mathbf{0}_{L\times K}$ denoting the $L\times K$ all-zero matrix. Next, we calculate the distances $\left\{ d_{l,k} \right\} _{l\in \mathcal{L},k\in \mathcal{K}}$ between all possible AP-user pairs. Based on these distances, we iteratively allocate $K$ out of $L$ APs to the $K$ users until each user is associated with a unique AP. Specifically, in the $j$-th iteration, let $\mathcal{L}^{\left( j-1 \right)}$ and $\mathcal{K}^{\left( j-1 \right)}$ denote the sets of unpaired AP and user indexes obtained from the $(j-1)$-th iteration, where we have $\left| \mathcal{L}^{\left( j-1 \right)} \right|=L-(j-1)$ and $\left| \mathcal{K}^{\left( j-1 \right)} \right|=K-(j-1)$. Then, we select the AP–user pair $(l^{\left( j \right)}, k^{\left( j \right)})$ that corresponds to the minimum $d_{l,k}$ among the unpaired AP and user sets: 
	
	\vspace{-0.2cm}
	\small
	\begin{equation}
		\left( l^{\left( j \right)},k^{\left( j \right)} \right) =\underset{l\in \mathcal{L}^{\left( j-1 \right)},k\in \mathcal{K}^{\left( j-1 \right)}}{\text{arg\ }\min}d_{l,k}. \label{12}
		\vspace{-0.2cm}
	\end{equation}\normalsize
	According to \eqref{12}, AP $l^{\left( j \right)}$ is allocated to user $k^{\left( j \right)}$. This allocation is recorded in matrix $\mathbf{B}$ as
	
	\vspace{-0.25cm}
	\small
	\begin{equation}
		b_{l^{\left( j \right)},k^{\left( j \right)}}=1\Rightarrow \mathbf{B}\left( l^{\left( j \right)},k^{\left( j \right)} \right) =1.
		\vspace{-0.2cm}
	\end{equation}\normalsize
	Thereafter, we remove the indexes of AP $l^{\left( j \right)}$ and user $k^{\left( j \right)}$ from $\mathcal{L}^{\left( j-1 \right)}$ and $\mathcal{K}^{\left( j-1 \right)}$, respectively, yielding 
	
	\vspace{-0.3cm}
	\small
	\begin{subequations}
		\begin{align}
			\mathcal{L}^{\left( j \right)}=\mathcal{L}^{\left( j-1 \right)}\setminus \left\{ l^{\left( j \right)} \right\} ,\\ \mathcal{K}^{\left( j \right)}=\mathcal{K}^{\left( j-1 \right)}\setminus \left\{ k^{\left( j \right)} \right\} .
		\end{align}
	\end{subequations}\normalsize
	Upon completion of $K$ iterations, the set of unpaired users becomes empty ($\mathcal{K}^{(K)} = \varnothing$), confirming that each user is served by exactly one AP in compliance with constraint \eqref{11c}.
	
	\textit{2) Remaining AP Assignment:} Since larger AP-user distances yield weaker links and offer limited gains for sum-rate maximization, it is preferable to associate users with nearby APs. Therefore, at this stage, we aim to assign each of the remaining $\left| \mathcal{L}^{\left( K \right)} \right|=L-K$ unpaired APs to its nearest user. For each unpaired AP, denoted by $l\in \mathcal{L}^{\left( K \right)}$, we find the user $k^{[l]}$ that is closest to AP $l$, i.e.,
	
	\vspace{-0.2cm}
	\small
	\begin{equation}
		k^{[l]}=\text{arg\ }\underset{k^{[l]}\in \mathcal{K}}{\min}\ d_{l,k},\ \ \ l\in \mathcal{L}^{\left( K \right)}.
		\vspace{-0.2cm}
	\end{equation}\normalsize
	AP $l$ is then allocated to user $k^{[l]}$, and the allocation is recorded in $\mathbf{B}$. The final output is the AP-user association matrix $ \mathbf{B} $, representing the overall AP–user associations.
	
	The proposed algorithm ensures that each user is served by at least one AP and that no AP remains idle. It should be noted that (P2) is an approximate surrogate of (P1), obtained by omitting the effects of small-scale fading and interference. Although this leads to a suboptimal solution, Algorithm 1 provides an effective AP-user association with low computational complexity, requiring only basic operations such as sorting and comparison. Specifically, the complexity order for solving problem (P2) is $\mathcal{O}\left( KL \right) $.
	\vspace{-3mm}
	
	\begin{algorithm}[!t]
		\small 
		\caption{Two-stage AP-user association strategy.}\label{alg:alg1}
		\begin{algorithmic}[1] 
			\renewcommand{\algorithmicrequire}{ \textbf{Input}}\REQUIRE $L,~K$, AP and user positions.
			\renewcommand{\algorithmicrequire}{ \textbf{Initialization}} \REQUIRE  $\mathcal{L}^{\left( 0 \right)}=\mathcal{L},\ \mathcal{K}^{\left( 0 \right)}=\mathcal{K}, \mathbf{B}=\mathbf{0}_{L\times K}. $
			\renewcommand{\algorithmicrequire}{ \textbf{Stage 1:}} \REQUIRE
			\STATE Calculate $d _{l,k} ,\forall l\in \mathcal{L},\forall k\in \mathcal{K}$.
			\FOR{$j=1$ to $K$}
			\STATE For $ l\in \mathcal{L}^{\left( j-1 \right)}, k\in \mathcal{K}^{\left( j-1 \right)}$, find the minimum $d_{l,k}$ and the corresponding AP-user pair $\left( l^{\left( j \right)},k^{\left( j \right)} \right)$.
			\STATE Allocate AP $l^{\left( j \right)}$ to user $k^{\left( j \right)}$, and update $\mathbf{B}$.
			\STATE Remove $l^{\left( j \right)}$ from $\mathcal{L}^{\left( j-1 \right)}$ and $k^{\left( j \right)}$ from $\mathcal{K}^{\left( j-1 \right)}$.
			\ENDFOR
			\renewcommand{\algorithmicrequire}{ \textbf{Stage 2:}} \REQUIRE
			\FOR{each AP $l \in \mathcal{L}^{\left( K \right)}$}
			\STATE Find the nearest user $k^{[l]}$.
			\STATE Allocate AP $l$ to user $k^{[l]}$, and update $\mathbf{B}$.
			\ENDFOR
			\renewcommand{\algorithmicrequire}{ \textbf{Output}}\REQUIRE AP-user association matrix $\mathbf{B}$.
		\end{algorithmic}
		\normalsize 
		\label{alg1}
	\end{algorithm}

	\subsection{RA Pointing Vector Optimization}
	After AP-user association, our objective is to optimize the RA pointing vectors at the APs. Accordingly, we simplify problem (P1) as follows:
	
	\vspace{-0.4cm}
	\small
	\begin{subequations}\label{16}
		\begin{align}
			\left( \text{P}3 \right) :\max\limits_{\mathcal{F}}&\,\,\sum_{k=1}^K{R_k}\label{16a}\\
			\text{s.t.} \;&\lVert \vec{\mathbf{f}}_l \rVert =1,\forall l. \label{16b}
		\end{align}
	\end{subequations}\normalsize
	As the achievable rate $R_k$ in~\eqref{16a} contains complex fractional structures, problem (P3) is challenging to solve directly. Therefore, we apply the quadratic transform to the SINR term in $R_k$, yielding the following equivalent problem:
	
	\vspace{-0.3cm}
	\small
	\begin{subequations}\label{17}
		\begin{align}
			\left( \text{P}4 \right) :\underset{\mathcal{F},\mathcal{Z}}{\max}&\sum_{k=1}^K{\log _2\left( 1+\tilde{\gamma}_k\left( \vec{\mathbf{f}}_l,z_k \right) \right)} \label{17a}\\ 
			\text{s.t.} \;&\lVert \vec{\mathbf{f}}_l \rVert =1,\forall l, \label{17b}
		\end{align}
	\end{subequations}\normalsize
	where $z_k$ is introduced by the quadratic transform for each user $k$, and $\mathcal{Z}=\left\{ z_1,...,z_K \right\} $ denotes the collection of $z_k$. The quadratic transform reformulates the SINR $\gamma_k$ in the objective function as 
	
	\vspace{-0.4cm}
	\small
	\begin{align}\label{18}
		\tilde{\gamma}_k\left( \vec{\mathbf{f}}_l,z_k \right) &=2\sqrt{P}~\mathrm{Re}\left\{ z_{k}^{*}\left( \sum_{l=1}^L{h_{l,k}\left( \vec{\mathbf{f}}_l \right) b_{l,k}\phi _{l,k}^{*}} \right) \right\}   \nonumber \\
		-\left| z_k \right|^2&\left( P\sum_{i\ne k}{\left| \sum_{l=1}^L{h_{l,k}\left( \vec{\mathbf{f}}_l \right) b_{l,i}\phi _{l,i}^{*}} \right|}^2+\sigma ^2 \right) ,
	\end{align}\normalsize
	where $\mathrm{Re}\left\{ \cdot \right\}$ denotes the real part of a complex number. Subsequently, we alternately optimize $\vec{\mathbf{f}}_l$ and $z_k$ to maximize the objective function~\eqref{17a}. For given $\vec{\mathbf{f}}_l$, the optimal $z_k$ has the following closed-form solution\cite{ref13}:
	
	\vspace{-0.2cm}
	\small
	\begin{equation}\label{19}
		z_{k}^{\star}=\frac{\sum_{l=1}^L{\sqrt{P}h_{l,k}\left( \vec{\mathbf{f}}_l \right) b_{l,k}\phi _{l,k}^{*}}}{P\sum_{i\ne k}{\left| \sum_{l=1}^L{h_{l,k}\left( \vec{\mathbf{f}}_l \right) b_{l,i}\phi _{l,i}^{*}} \right|}^2+\sigma ^2}.
	\end{equation}\normalsize
	
	Next, we focus on optimizing $\vec{\mathbf{f}}_l$ for given $z_k$. As $G_{l,k}\left( \vec{\mathbf{f}}_l \right)$ in $h_{l,k}$ is a piecewise function, objective function~\eqref{17a} remains non-convex. To address this issue, we approximate $G_{l,k}\left( \vec{\mathbf{f}}_l \right)$ using the softplus function, which provides a smooth and differentiable surrogate as
	
	\vspace{-0.2cm}
	\small
	\begin{equation}
		G_{l,k}^{'}\left( \vec{\mathbf{f}}_l \right) =G_0\left[ \frac{\ln \left( 1+e^{m\vec{\mathbf{f}}_{l}^{T}\vec{\boldsymbol{q}}_{l,k}} \right)}{m} \right] ^{2p},
	\end{equation}\normalsize
	where $m$ controls the smoothness of the approximation. In addition, we apply successive convex approximation (SCA), iteratively approximating \eqref{17a} with a convex surrogate. Without loss of generality, we present the process of the $(i+1)$-th iteration, where the values of $\vec{\mathbf{f}}_{l}$ obtained in the $i$-th iteration are denoted by $\vec{\mathbf{f}}_{l}^{\left( i \right)}$. By using the first-order Taylor expansion at $\vec{\mathbf{f}}_l$, $
	h_{l,k}\left( \vec{\mathbf{f}}_l \right)$ can be linearized as $\Lambda _{l,k}^{\left( i+1 \right)} $, and objective~\eqref{17a} can be established as $U\left( \vec{\mathbf{f}}_{l} \right)$, both of which are shown at the top of next page. 
	\newcounter{TempEqCnt}                   
	\setcounter{TempEqCnt}{\value{equation}} 
	\setcounter{equation}{19}                
	\begin{figure*}[h]
		\vspace{-5mm}

		\vspace{-0.2cm}
		\small
		\begin{equation}
			\begin{aligned}
				\Lambda _{l,k}^{\left( i+1 \right)}\left( \vec{\mathbf{f}}_l \right) \triangleq h_{l,k}\left( \vec{\mathbf{f}}_{l}^{\left( i \right)} \right) +p\sqrt{\frac{\beta \left( d_{l,k} \right) G_0\kappa}{\kappa +1}}e^{-j\frac{2\pi}{\lambda}d_{l,k}}\left( \frac{\ln \left( 1+e^{m\vec{\mathbf{f}}_{l}^{\left( i \right) ^T}\vec{\boldsymbol{q}}_{l,k}} \right)}{m} \right) ^{p-1}\frac{e^{m\vec{\mathbf{f}}_{l}^{\left( i \right) ^T}\vec{\boldsymbol{q}}_{l,k}}}{1+e^{m\vec{\mathbf{f}}_{l}^{\left( i \right) ^T}\vec{\boldsymbol{q}}_{l,k}}}\left( \vec{\mathbf{f}}_l-\vec{\mathbf{f}}_{l}^{\left( i \right)} \right) ^T\vec{\boldsymbol{q}}_{l,k}.
			\end{aligned}
		\end{equation}\normalsize
		\setcounter{equation}{\value{TempEqCnt}} 
		\setcounter{equation}{21}        		
		
		\vspace{-4mm}
		
		\newcounter{TempEqCnt1}                   
		\setcounter{TempEqCnt1}{\value{equation}} 
		\setcounter{equation}{20}                
		
		\vspace{-0.2cm}
		\small
		\begin{equation} \label{22}
			\begin{aligned}
				U\left( \vec{\mathbf{f}}_l \right) \triangleq \sum_{k=1}^K{\log _2\left( 1+2\sqrt{P}\text{Re}\left\{ z_{k}^{*}\left( \sum_{l=1}^L{\Lambda _{l,k}^{\left( i+1 \right)}\left( \vec{\mathbf{f}}_l \right) b_{l,k}\phi _{l,k}^{*}} \right) \right\} -\left| z_k \right|^2\left( P\sum_{i\ne k}{\left| \sum_{l=1}^L{\Lambda _{l,k}^{\left( i+1 \right)}\left( \vec{\mathbf{f}}_l \right) b_{l,i}\phi _{l,i}^{*}} \right|^2}+\sigma ^2 \right) \right)}.
			\end{aligned}
		\end{equation}\normalsize
	\hrulefill
	\end{figure*}
	\setcounter{equation}{21}                
	
	Thus, problem (P4) can be transformed to
	\setcounter{equation}{\value{TempEqCnt1}} 
	
	\vspace{-0.3cm}
	\small
	\begin{subequations}
		\begin{align}
			\left( \text{P}5 \right) :\underset{\mathcal{F} }{\max}\ & U\left( \vec{\mathbf{f}}_{l} \right)  \\
			\text{s.t.} \;&~\eqref{16b}.
		\end{align}
	\end{subequations}\normalsize
	However, problem (P5) remains non-convex due to the unit-norm constraint on $\vec{\mathbf{f}}_l$. To address this, we relax the equality constraint~\eqref{16b} as $\lVert \vec{\mathbf{f}}_l \rVert \le 1$, yielding the following problem,
	
	\vspace{-0.2cm}
	\small
	\begin{subequations} 
		\begin{align}
			\left( \text{P}6 \right) :\max\limits_{\mathcal{F}}\ & U\left( \vec{\mathbf{f}}_{l} \right)\label{24a}\\
			\text{s.t.} \;&\lVert \vec{\mathbf{f}}_l \rVert \le 1,\forall l.
		\end{align}
	\end{subequations}\normalsize
	
	It can be verified that problem (P6) is convex and can be solved by the CVX solver\cite{ref14}. Note that the optimal value obtained by problem (P6) serves as an upper bound for that of problem (P5) due to the relaxation of the equality constraint ~\eqref{16b}. After obtaining the optimal solution of $\vec{\mathbf{f}}_{l}$ through Algorithm 2, the pointing vector needs to be recovered as a unit vector, i.e., $\vec{\mathbf{f}}_{l}^{\star}=\frac{\vec{\mathbf{f}}_l}{\lVert \vec{\mathbf{f}}_l \rVert}$. This normalization only scales $\vec{\mathbf{f}}_l$ to unit length without altering its direction.
	
	We summarize the proposed fractional programming-based algorithm in Algorithm 2. Since the optimal objective function is non-decreasing over the iterations, Algorithm 2 is guaranteed to converge. In terms of computational complexity, the cost of updating $z_{k}$ is negligible, while optimizing $\vec{\mathbf{f}}_{l}$ in (P6) incurs a complexity of $\mathcal{O}\left( L^{3.5}\ln \left( 1/\varepsilon \right) \right) $ per iteration, where $\varepsilon $ denotes the accuracy threshold for convergence. Hence, the overall complexity of Algorithm 2 is  $\mathcal{O}\left( IL^{3.5}\ln \left( 1/\varepsilon \right) \right) $, where $I$ represents the number of iterations required for convergence.
	
	\begin{algorithm}[!t]
		\small 
		\caption{Fractional Programming-based Algorithm.}\label{alg:alg2}
		\begin{algorithmic}[0] 
			\REQUIRE  Pointing vector $\vec{\mathbf{f}}_{l}^{\left( 0 \right)}$ and threshold $\xi >0 $.
			\STATE Initialization: $i\leftarrow0$.
			\REPEAT 
			\STATE Given $\vec{\mathbf{f}}_{l}^{\left( i \right)}$, calculate $z_{k}^{(i+1)}$ according to \eqref{19}.
			\STATE Given $z_{k}^{(i+1)}$ and $\vec{\mathbf{f}}_{l}^{\left( i \right)}$, obtain $\vec{\mathbf{f}}_{l}^{\left( i+1 \right)}$ by solving problem (P6).
			\STATE Update $i=i+1$.
			\UNTIL The fractional increase of the sum rate in \eqref{17a} falls below threshold $\xi$ or iteration number $i$ reaches the pre-designed number of iterations $I$.
			\ENSURE $\vec{\mathbf{f}}_{l}=\vec{\mathbf{f}}_{l}^{\left( i \right)}$ and $z_{k}=z_{k}^{(i)}$.
		\end{algorithmic}
		\normalsize
	\end{algorithm}
	
	\section{Simulation Results}
	In this section, we evaluate an RA-enhanced downlink cell-free system employing the proposed AP-user association strategy and fractional programming-based algorithm. We consider $L$ APs and $K$ users randomly distributed in a $300\  \text{m}\times300\ \text{m}$ area. The system operates at 2.4 GHz ($\lambda=0.125$ m) with noise power $\sigma^2=-94$ dBm. The large-scale channel power gain is modeled as $ \beta\left( d_{l,k} \right) =C_0\left( d_0/d_{l,k} \right) ^{\alpha}$, with a reference power gain $C_0=-40$ dB at $d_0 = 1$ m and a path loss exponent of $\alpha=2.3$. Unless otherwise stated, parameters are set as $p=6$, $P=24$ dBm, $\kappa=7.94$, and $m=20$ (as in \cite{ref15}). Results are averaged over 100 Monte Carlo realizations and compared against the following benchmark schemes: 
	\begin{itemize}
		\item \textbf{Fixed directional antenna-based scheme}: In this scheme, the pointing vectors are fixed as $\vec{\mathbf{f}}_l=\left[ 1,0,0 \right]^{T} ,\forall l$.
		\item \textbf{Isotropic antenna-based scheme}: In this scheme, each AP serves all users, the directional gain in~\eqref{1} is replaced by $G_e\left( \epsilon ,\varphi \right) =G_0\cos ^{2p}\left( \epsilon \right)$, with $G_0=1$ and $p=0$, and a minimum mean-squared error (MMSE) precoding scheme is adopted.
		\item \textbf{RA-user alignment scheme}: In this scheme, each RA steers its boresight directly toward its associated user.
	\end{itemize}
	
	\vspace{-0.5cm}
	\begin{figure}[!h]	
		\centering
		\includegraphics[width=0.7\linewidth]{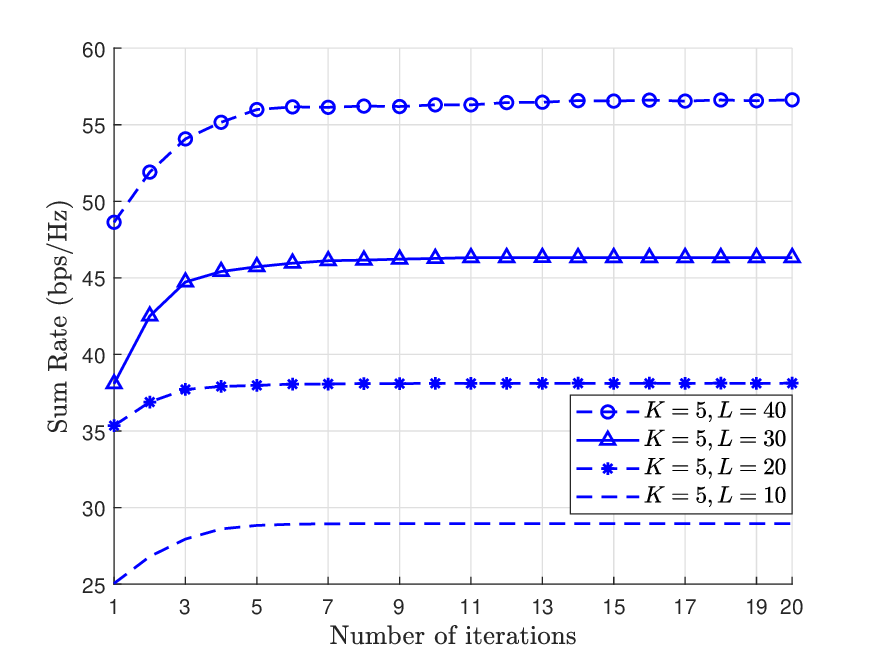}
		\vspace{-0.3cm}
		\caption{\centering{Convergence behavior of the proposed algorithm.}}
		\label{fig_3}
		\vspace{-0.2cm}
	\end{figure}	
	\vspace{-0.0cm}

	\begin{figure*}[t]
		\vspace{-0.6cm}
		\centering
		
		\begin{minipage}[t]{0.24\textwidth}
			\centering
			\includegraphics[width=\linewidth]{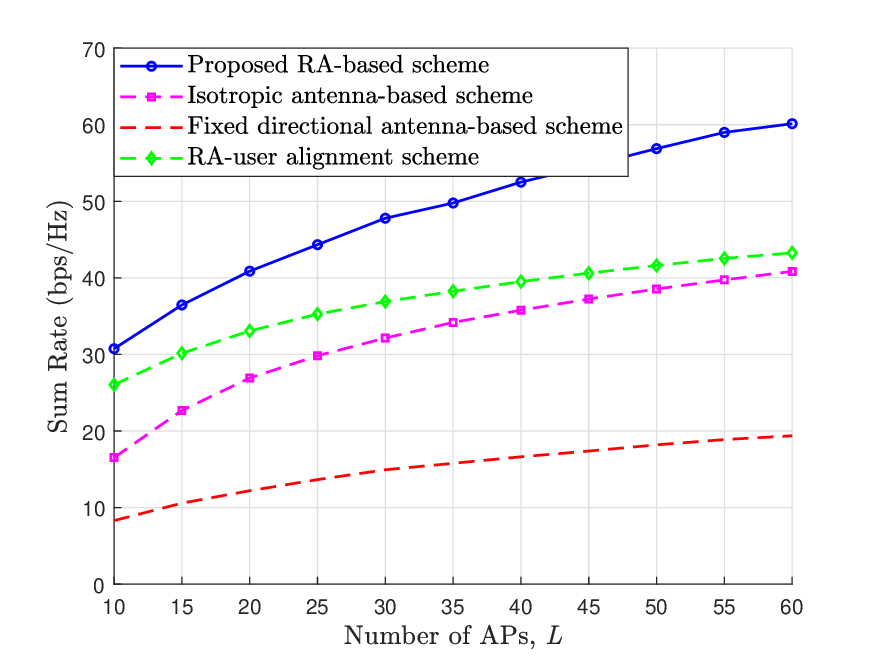}
			\vspace{-0.5cm}
			\caption{Sum rate (bps/Hz) vs. the number of APs for $K=5$.}
			\label{fig_4}
		\end{minipage}
		\hfill
		\begin{minipage}[t]{0.24\textwidth}
			\centering
			\includegraphics[width=\linewidth]{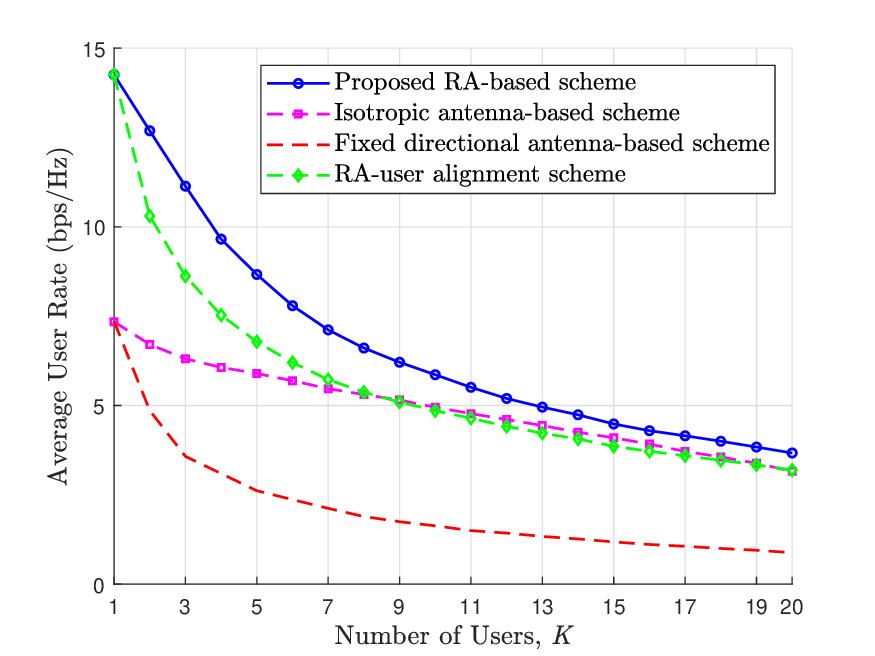}
			\vspace{-0.5cm}
			\caption{Average per-user rate (bps/Hz) versus the number of users for $L=30$.}
			\label{fig_5}
		\end{minipage}
		\hfill
		\begin{minipage}[t]{0.24\textwidth}
			\centering
			\includegraphics[width=\linewidth]{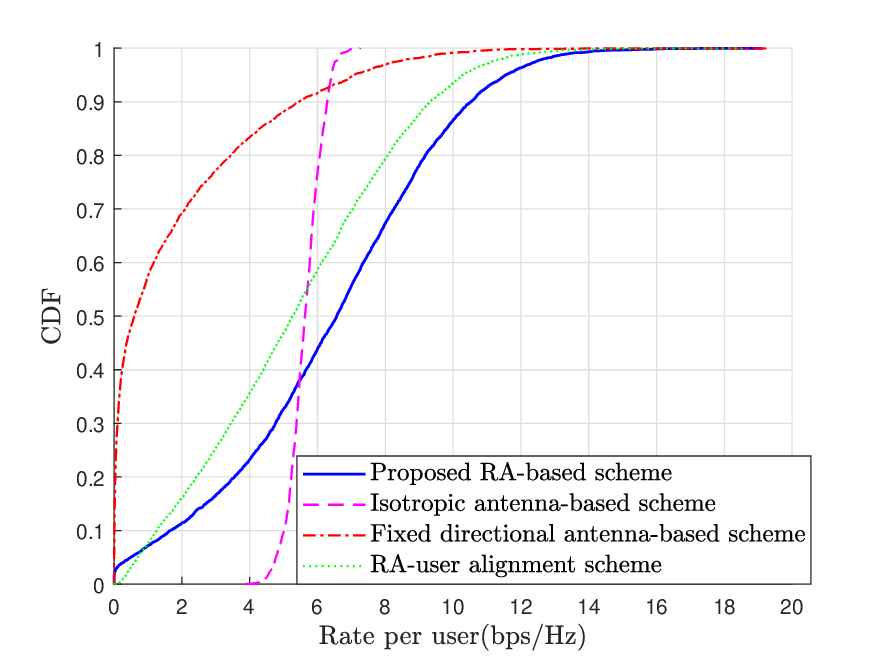}
			\vspace{-0.5cm}
			\caption{CDF of rate per user, where $K=10$ and $L=30$.}
			\label{fig_6}
		\end{minipage}
		\hfill
		\begin{minipage}[t]{0.24\textwidth}
			\centering
			\includegraphics[width=\linewidth]{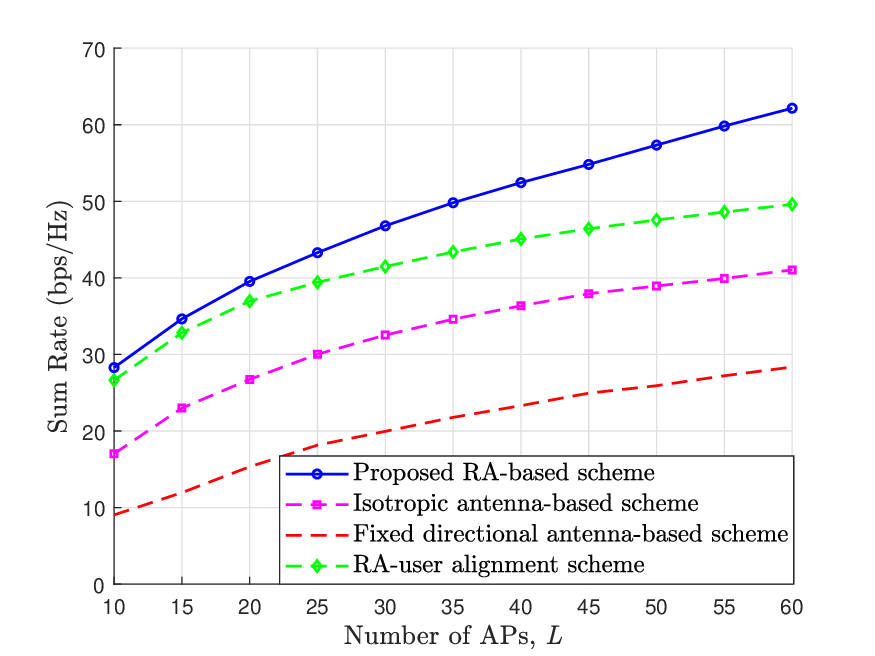}
			\vspace{-0.5cm}
			\caption{Sum rate (bps/Hz) vs. the number of APs for $K=5$ (each AP serves multiple users).}
			\label{fig_7}
		\end{minipage}
		
		\vspace{0cm}
	\end{figure*}
	
	In Fig. 3, we illustrate the convergence behavior of the proposed algorithms. As expected, for all considered numbers of APs $L$, the sum rate increases monotonically with the number of iterations. Although an increased number of APs results in marginally slower convergence, all configurations converge within 10 iterations, verifying the algorithm’s effectiveness.
	
	Fig. 4 illustrates the sum rate performance versus the number of APs $L$ for a fixed number of users $K=5$. As expected, the sum rate increases with $L$ for all considered schemes.  Moreover, the proposed RA-based scheme and the RA-user alignment scheme outperform other benchmark schemes in terms of the sum rate performance. This performance gain is mainly due to the capability of RAs to dynamically steer their boresight directions toward the desired users, thereby focusing the radiation power for SINR enhancement. However, as $L$ increases, the RA-user alignment scheme approaches the isotropic antenna-based scheme, indicating that simply steering each RA toward its served user overlooks inter-user interference and thus yields only limited performance improvement. Notably, the proposed RA-based scheme achieves an additional performance gain over the RA-user alignment baseline. This underscores the effectiveness of the proposed joint optimization algorithm, which also accounts for the impact of inter-user interference rather than merely pursuing the maximization of the associated user's signal power.
	
	
	Fig. 5 illustrates the average per-user rate versus the number of users $K$ for different schemes. As $K$ increases, the average per-user rate decreases for all considered schemes since more users will lead to more severe multiuser interference. Notably, the proposed RA-based scheme consistently achieves the best performance and demonstrates superior robustness against inter-user interference. This improvement is primarily attributed to the ability of RAs to adaptively rotate their boresight directions, thereby concentrating signal power toward intended users and reducing the interference caused by signal energy leakage to other users. Meanwhile, as $K$ increases for a fixed number of APs, the impact of multiuser interference becomes increasingly pronounced. Moreover, users are more crowded in the angular domain, which reduces the interference-suppression flexibility offered by boresight rotation. As a result, the relative performance gain of the proposed RA-based scheme gradually decreases. When $K$ is large, multiuser interference dominates, and the isotropic antenna-based scheme with MMSE precoding can even outperform the RA-user alignment scheme, as the latter does not explicitly account for inter-user interference. Additionally, the fixed directional antenna-based scheme exhibits poor performance, as its inability to adjust boresight directions leads to limited coverage and poor interference management. This observation highlights the importance of directional flexibility in enhancing overall communication quality.
	
	Fig. 6 shows the cumulative distribution function (CDF) of the rate per user for the considered schemes. As compared to the RA-user alignment scheme, for the proposed RA-based scheme, a greater proportion of users achieve high data rates, indicating that optimized boresight control significantly improves per-user performance. In addition, the fixed directional antenna-based scheme results in zero-rate transmission for a subset of users, primarily due to its inability to adapt the boresight directions, which leads to coverage blind spots.
	
	
	Fig. 7 illustrates the potential of the AP serving multiple users. In this scenario, after applying the proposed antenna orientation optimization algorithm, each AP adopts the MMSE precoding to simultaneously serve all users within its coverage region. Notably, when each AP serves multiple users with proper precoding, all the considered schemes exhibit improved performance, indicating that the RA-enabled multi-user transmission paradigm holds significant potential for further performance enhancement. In addition, as the number of APs increases, all schemes benefit from macro-diversity and coordinated transmission. However, the proposed RA-based scheme achieves a more pronounced sum-rate improvement, since RA can exploit the directivity of the antenna and can enhance desired links through boresight adaptation while reducing interference leakage toward unintended users.

	\section{Conclusion}
	In this letter, we proposed a novel RA-enhanced cell-free system, where the RA pointing vectors can be adjusted to change the directional gain pattern for maximizing the system sum rate. Specifically, we developed a two-stage strategy to solve the AP-user association problem and a fractional programming-based algorithm to obtain a high-quality solution for the RA pointing vectors. Simulation results demonstrate that, by optimizing each RA’s pointing vector, the RA-equipped APs significantly enhance the directional gain toward intended users while suppressing interference in undesired directions, thereby achieving a higher system sum rate than various benchmark schemes. Future research may consider more general scenarios where a single AP serves multiple users simultaneously, as well as the integration of RA with orthogonal frequency division multiplexing (OFDM).
	
	\vspace{-2mm}
	\bibliographystyle{IEEEtran}
	
	\vfill
	
\end{document}